\documentclass[12pt]{aastex62}
\usepackage{lineno}

\usepackage{color}
\usepackage{graphicx}
\usepackage{longtable}
\usepackage{multirow}
\usepackage{url}
\usepackage{subfigure}
\usepackage{amsmath}
\usepackage{lipsum}
\usepackage{threeparttable}

\begin{document}
	
\title{Repeating fast radio bursts reveal memory from minutes to an hour}

    \author[0000-0003-4157-7714]{F. Y. Wang} \affiliation{School of Astronomy and Space Science, Nanjing
	University, Nanjing 210093, China} \affiliation{Key Laboratory of
	Modern Astronomy and Astrophysics, Nanjing University, Nanjing 210093, China}
   \affiliation{Purple Mountain Observatory, Chinese Academy of Sciences, Nanjing, 210023, China}
	
	\author[0000-0001-6021-5933]{Q. Wu}
	\affiliation{School of Astronomy and Space Science, Nanjing University, Nanjing 210093, China}
	
	\author[0000-0002-7835-8585]{Z. G. Dai} \affiliation{Department of Astronomy, University of Science and Technology of China, Hefei 230026, China}

	\correspondingauthor{F. Y. Wang}
	\email{fayinwang@nju.edu.cn}
	
\begin{abstract}
Fast radio bursts (FRBs) are brief, luminous pulses with unknown physical origin. The repetition pattern of FRBs contains essential information about their physical nature and emission mechanisms. 
Using the two largest samples of FRB 20121102 and FRB 20201124A,
we report that the sources of the two FRBs reveal memory over a large range of timescales, from a few minutes to about an hour.
The memory is detected from the coherent growths in
burst-rate structures and the Hurst exponent.
The waiting time distribution displays an approximate power-law tail, which is consistent with a Poisson model with a time-varying rate. 
From cellular automaton simulations, we find that these
characteristics can be well understood within the physical framework of a self-organized criticality 
system driven in a correlation way, such as random walk functions. These properties indicate that the triggers of bursts are correlated, preferring the crustal failure mechanism of neutron stars. 
\end{abstract}
	

\section{Introduction}
Fast radio bursts (FRBs) are bright millisecond-duration astronomical transients \citep{Lorimer2007,Xiao2021,Zhang2022,Petroff2022}. Some FRBs, such as FRB 20121102 \citep{Spitler2016} and FRB 20201124A \citep{Lanman2022}, are known to repeat, but it remains
unclear whether repeating FRBs are prevalent or uncommon
sources. Observations show that the activity of repeating FRBs is intermittent \citep{Spitler2016}. Bursts may be
separated by a few milliseconds to a few thousand seconds during active periods, while, no burst is detected in a long-time campaign. It has been shown that FRB 20180916B has a 16.35-day periodicity in
its burst times \citep{Chime/FrbCollaboration2020}, and FRB 20121102 also shows a possible periodic activity \citep{Rajwade2020}, which tightly constrain the physical models of FRBs \citep{Platts2019}. 
Therefore, the repetition pattern of FRBs offers an important clue to the
nature of the progenitor \citep{Zhang2020}, emission mechanism \citep{Li2021}, and local environment \citep{Petroff2022,Wang2022}.

Information about FRB central engines and emission mechanisms
can be obtained from their waiting times $\Delta t$ (the time between consecutive bursts). For a relatively small sample of FRB 20121102, it has been found that the waiting times deviate from the Poisson distribution \citep{Wang2017,Oppermann2018,Wadiasingh2019,Cheng2020}.
The waiting times have been found to follow a bimodal distribution with peaks at milliseconds and tens of seconds from the large samples of FRB 20121102 \citep{Li2021,Jahns2023,Hewitt2021} and FRB 20201124A \citep{Xu2022}. 
It is possible that the short waiting time is only the substructure of longer bursts. 
The dispersion measure (DM) varied significantly between different bursts and the substructure of bursts would have the same DM. 
Two log-normal functions approximated to the Poisson model can describe the bimodal waiting time distribution \citep{Li2021,Xu2022}, which indicates that the central engine emits FRBs in a stochastic process \citep{Katz2023}. 
Monte Carlo simulations also showed that the observed log-normal distributions of
waiting times require that the central engine emits
FRBs randomly \citep{Li2021}. However, a single Poisson or Weibull distribution
cannot fit the whole waiting time distribution \citep{Zhang2018,Gourdji2019,Cruces2021,Aggarwal2021,Zhang2021}, especially for the waiting times with $\Delta t< 0.1$ s. 
In this study, we analyzed the two largest FRB samples of FRB 20121102 \citep{Li2021} and FRB 20201124A \citep{Xu2022} detected by the Five-hundred-meter Aperture Spherical radio Telescope (FAST), 
which consist of 1659 and 2610 bursts, respectively.
The stationary Poisson process and non-stationary Poisson process are used to fit them separately. 

For non-stationary Poisson processes, the events have a certain amount of ``memory" \citep{Aschwanden2021}, which is characterized by the coherent growth in
burst-rate structures $\lambda(t)$. These time structures $\lambda(t)$ represent partial time segments and can be extracted from the
entire time profile in the total time interval of the observed burst rates. Usually, it shows a polynomial function, $\lambda (t)\propto t^p$, which can be used to distinguish between linear ($p = 1$) and nonlinear ($\bf p\ne 1$) time evolutions.
Besides, memory effects can be detected using the self-similarity Hurst exponent \citep{Hurst1951}. The Hurst exponent $H$ is a measure of the secular memory of a time series. It has been widely used in the study of many natural phenomena, such as geology \citep{Barani2018}, geomagnetism \citep{Wanliss2003}, solar activity \citep{Oliver1996,Lepreti2000,Zhou2014} and so on.
A value of $H=0.5$
indicates that the behavior of the time series corresponds to a stochastic process with no memory. However, 
$0.5<H\leq 1$ means that the memory is persistent, supporting that a large event is followed by a larger one in the future.

In this letter, we show that repeating FRBs show memory from minutes to an hour.  The structure of this letter is as follows.
The waiting time distributions of repeating FRBs are presented in Section \ref{sec:non-stationary}. 
In Section \ref{sec:memory}, we search for the ``memory" of repeating FRBs in two different ways.
The cellular automation simulation is conducted in Section \ref{sec:simulation}. 
In the last, the discussion and conclusions are given in Section \ref{sec:discussion}.

\section{Waiting time distributions for repeating FRBs} \label{sec:non-stationary}
Here we analyze the waiting time distributions for repeating FRB 20121102 and FRB 20201124A observed by FAST \citep{Li2021, Xu2022}. 
Here we combine the data of FRB 20201124A detected by FAST from April to May \citep{Xu2022} and September in 2021 \citep{Zhou2022}. 
The observations of FRB 20121102 detected by Arecibo in 2016 \citep{Hewitt2022} and 2018 \citep{Jahns2023} are not included, by considering the instrument effects between different telescopes. 

\begin{figure*}
	\includegraphics[width=\textwidth]{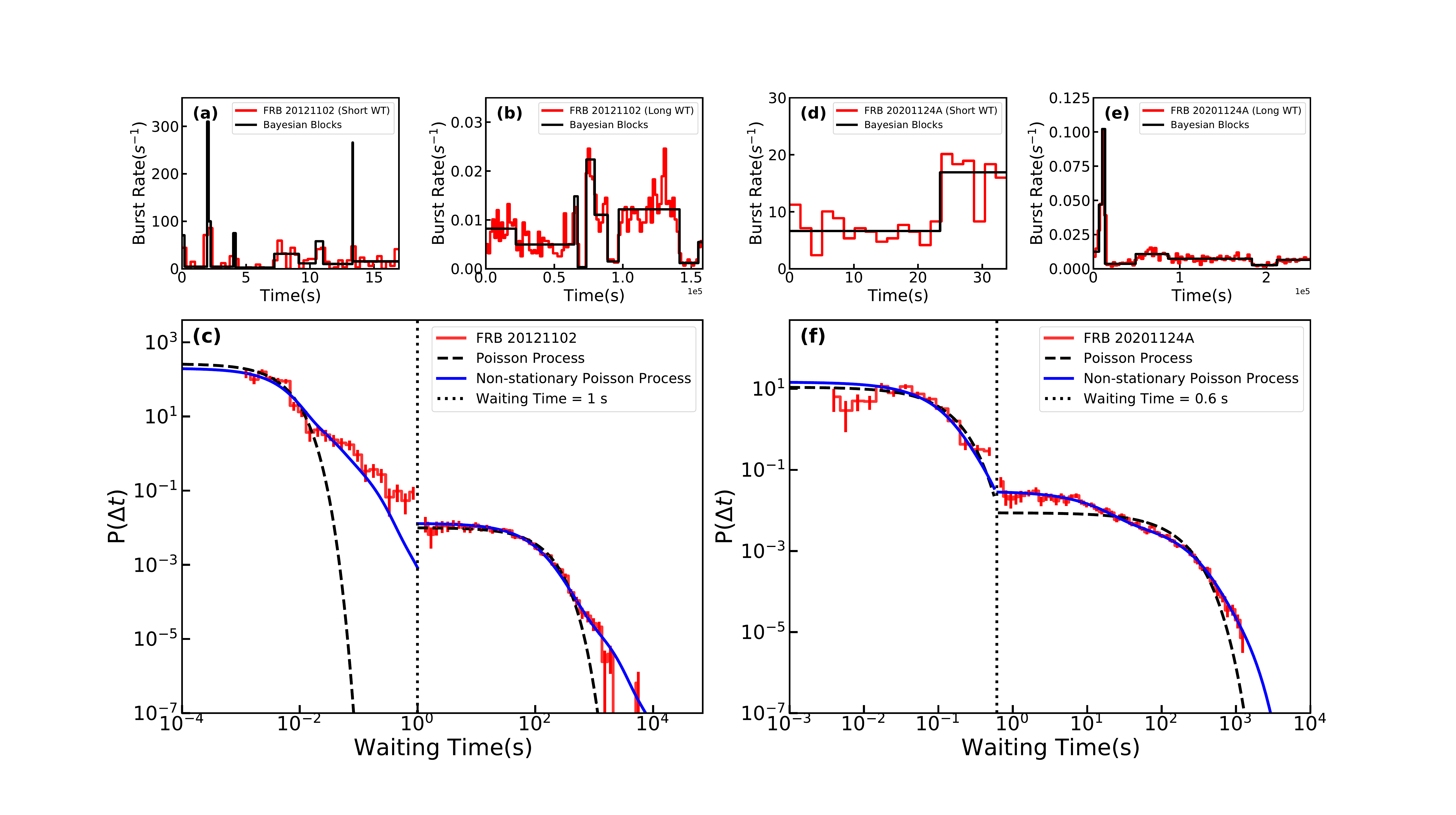}
	\caption{\textbf{The waiting time distribution of repeating FRBs.} Panel (a). Bayesian blocks decomposition of the burst rate for short waiting times ($\Delta t<1$ s) of FRB 20121102.
		Panel (b). Bayesian blocks decomposition of the burst rate for long waiting times ($\Delta t>1$ s) of FRB 20121102. Panel (c). The red stepwise line is the observed distribution of waiting time of FRB 20121102. The solid lines show the
		fitting result for a non-stationary Poisson process with a distribution of rates estimated
		from the data (panels a and b). The dashed lines give the fitting results of a Poisson process (equation (1)). From the reduced $\chi^2$, we can see that the distribution of long waiting times can be explained by a non-stationary Poisson process. Panel (d). Same as panel (a), but for FRB 20201124A. Panel (e). Same as Panel (b), but for FRB 20201124A. Panel (f). Same as panel (c), but for FRB 20201124A. The distribution of long waiting times for FRB 20201124A can be well explained by a  non-stationary Poisson process.}
	\label{fig:WTD}
\end{figure*}

The waiting times are logarithmically divided into several bins. The differential distribution of waiting times is defined as the fraction of bursts in each bin divided by the bin width $\Delta t_i$:
\begin{equation}
P(\Delta t_i)  = \frac{N_{\rm bin,i}}{N_{\rm total} \Delta t_i},
\end{equation}
where $N_{\rm bin,i}$ is the number of bursts per bin and $N_{\rm total}$ is the total number of bursts. The expected uncertainty of the differential distribution is
\begin{equation}
\sigma_{i} = \sqrt{\frac{{P(\Delta t_i)}}{\Delta t_i}}.
\end{equation}
The differential distribution of the waiting times $P(\Delta t)$ of FRB 20121102 and FRB 20201124A are shown as red stepwise lines in panels (c) and (f) of Figure \ref{fig:WTD}.
They both show two distinct components, e.g., short waiting times ($\Delta t<1$ s) and long waiting times ($\Delta t>1$ s). 
Two Poisson models are used to fit them separately.
If the bursts occur randomly with a constant rate $\lambda$, the waiting time distribution is given by 
\begin{equation}\label{Poisson}
P(\Delta t)=\lambda \exp(-\lambda \Delta t).
\end{equation}
We fit the waiting time distribution of repeating FRBs with a Poisson process (equation \ref{Poisson}), in which the burst rate is constant. The Monte Carlo Markov Chain (MCMC) analysis is used in fitting. The fitting results are shown as black \textbf{dashed} lines in panels (c) and (f) of Figure \ref{fig:WTD}. For FRB 20121102, the burst rates of the Poisson process are found to be 263.61 $\rm s^{-1}$ and 0.01 $\rm s^{-1}$ for the short waiting times and the long waiting times, respectively. For FRB 20201124A, the fitted burst rates of the Poisson process are 11.02 $\rm s^{-1}$ and 0.009 $\rm s^{-1}$ for the short waiting times and the long waiting times, respectively.

It has been found that the waiting times in a single observation can be described by a Poisson process if the short waiting times are excluded \citep{Cruces2021}.
Moreover, the burst rate varies between different observations significantly \citep{Li2021,Xu2022}. Therefore, we try to model the waiting time distribution by a non-stationary (time-dependent) Poisson process. 
The key issue is how to determine the time-dependent burst rate $\lambda (t)$.
The Bayesian block procedure is used to determine the mean burst rate as a function of time \citep{Scargle1998}. 
This method takes a sequence of times of bursts and determines a decomposition into a time block, in which the observed burst
occurrence can be regarded as a constant rate.
These blocks are characterized by a rate $\lambda_i$ and a duration $t_i$.
For non-stationary Poisson processes, the waiting time distribution of a piecewise-constant Poisson process with rates
$\lambda_i$ and intervals $t_i$ can be approximated as the superposition of multiple
exponential distributions \citep{Wheatland1998},
\begin{equation}\label{equ:nonPoisson}
P(\Delta t)=\Sigma_i \phi_i \lambda_i \exp(-\lambda_i \Delta t),
\end{equation}
where $\phi_i=\lambda_i t_i/\Sigma_i \lambda_i t_i$ is the fraction of bursts in a block.
The Bayesian Blocks algorithm divides the whole time series into several blocks by giving several change points. In each block, the event rate can be approximately regarded as a constant. 
Here we only consider the time interval of two consecutive bursts as waiting time since the observations of FRBs are not always continuous. Then we divide the waiting times of FRB 20121102 and FRB 20201124A into short waiting times and long waiting times according to the classification criteria of 1 s and 0.6 s, respectively. These criteria are based on the lowest valley between the two peaks of the bimodal distribution of waiting times.
After the classification, we splice the waiting time into a continuous time series and block it using the \texttt{astropy.stats.bayesian\_blocks} package of Python. We set $p_0 = 0.01$, which gives the false alarm probability to compute the prior. 
Panels (a) and (b) of Figure \ref{fig:WTD} shows the blocks for FRB 20121102, while panels (d) and (e) of Figure \ref{fig:WTD} present the blocks of FRB 20201124A.
The red line is the time point associated with the waiting time, ignoring the time without observation.

We can see that the waiting time distributions show power-law tails in Figure \ref{fig:WTD}, which can be explained by the non-stationary Poisson process.
The superposition of multiple exponential distributions (equation \ref{equ:nonPoisson}) 
with rate and duration derived from the Bayesian Blocks method can fit the waiting time distributions, especially for long waiting times. The fitting results are shown as blue solid lines in panels (c) and (f) of Figure \ref{fig:WTD}.

There are significant deviations from the fittings of the Poisson process to the observed waiting time distribution of FRB 20121102 and FRB 20201124A from Figure \ref{fig:WTD}. 
To quantitatively compare the fittings of the Poisson process and the non-stationary Poisson process respectively, we calculate the goodness-to-fit $\chi^2$-criterion (reduced $\chi^2$) \citep{Aschwanden2015}
\begin{equation}
\chi^2=\frac{1}{\left(n_{\rm bins}-n_{\rm{par}}\right)} \sum_{i=1}^{n_{\rm bins}} \frac{\left[P_{\rm{fit}}\left(\Delta t_{i}\right)-P_{\rm{obs}}\left(\Delta t_{i}\right)\right]^{2}}{\sigma_{i}^{2}},
\end{equation}
where $P_{\rm{fit}}(\Delta t_{i})$ is the fitted value, $P_{\rm{obs}}(\Delta t_{i})$ is the observed value, $n_{\rm bins}$ is the number of bins and $n_{\rm par}$ is the parameters of the model (here $n_{\rm par} = 1$). From the definition of $\chi^2$, the expected value of $\chi^2 \approx 1$ indicates that the fitted distribution is closest to the true distribution. The fitted $\chi^2$ for FRB 20121102 and FRB 20201124A are shown in Table \ref{chi2}. 

For the fitting results of the non-stationary Poisson process of short waiting times, the reduced $\chi^2$ are 3.12 and 6.22 for FRB 20121102 and FRB 20201124A, respectively. 
There is some discrepancy between the
observed and model distributions, e.g., there are too few bursts with observed waiting times below $10$ ms and too many bursts with observed waiting times above 0.1 seconds. 
The observational determination of short waiting time has not been unified. 
First, short waiting times are likely to be missed as a result of the overlap
of bursts close in time. Second, the high sensitivity of FAST can
resolve more sub-bursts. However, the definition of sub-burst is ambiguous at present \citep{Li2021,Jahns2023,Xu2022}. 
It has been found that the short waiting times of FRBs are contaminated by sub-bursts \citep{Jahns2023}.
From the above analysis, we have found that the short waiting times cannot be fitted by the Poisson model and the non-stationary Poisson model.
It may indicate a characteristic timescale of the central engine and the emission mechanism.

Clearly, it is a good qualitative agreement between the observed waiting time distributions and the non-stationary Poisson process for long waiting times ($\Delta t> 1$ s). The reduced $\chi^2$ of long waiting times are
$1.05$ and $1.35$ for FRB 20121102 and FRB 20201124A, respectively. 
In particular, the model distribution produces power-law-like
behavior after a nearly constant phase. 
The good qualitative agreement
of the non-stationary Poisson model and observed distributions for long waiting times provide strong
evidence that the repeating FRBs occur as a non-stationary Poisson process, rather than producing bursts in completely stochastic order.

\section{``Memory" of Repeating FRBs} \label{sec:memory}
``Memory" exists in a nonlinear process that manifests as a monotonic increase in event rate over time  \citep{Aschwanden2021}. 
Here we propose two independent ways to find the ``Memory" of repeating FRBs.

\subsection{The coherent growths in burst-rate structures}

We use this polynomial model of the burst rate to measure the timescale of memory for repeating FRBs.
They are selected as follows. First, a local burst rate peak $\lambda(t=t_{\rm max})$ is selected.
The peak is defined as $\lambda(t_{i-1})<\lambda(t_i)=\lambda(t=t_{\rm max})>\lambda(t_{i+1})$. Second, 
the absolute minimum $t_{\rm min}$ between two
subsequent peaks should require $\lambda(t_{\rm max, j-1})>\lambda(t=t_{\rm min})<\lambda(t_{\rm max, j+1})$.
The time structure between a minimum at $t_{\rm min}$ and a maximum
at $t_{\rm max}$ is characterized by a monotonic increase in the burst rate.
The time structures are sampled using an automated detection
algorithm for 3000 different time resolutions $\delta t$ in the range [1 s, 3000 s], linearly
spaced with $n_{\rm bin} = \lceil T_{\rm tot} / \delta t \rceil$, where $T_{\rm tot}$ is the total observational time from the first burst to the last burst excluding the unobserved time. 
The total number of detected time structures is 92, ranging from time resolutions of [32 s, 2236 s] for FRB 20121102 (Table \ref{data:121102}). 
The total number of detected time structures is 183, ranging from time resolutions of [16 s, 1514 s] for FRB 20201124A (Table \ref{data:201124}).
The time structure $\lambda(t)$ is fitted to the observed data by the standard least-squares optimization algorithm. 

For a series of burst arrival times, we calculate the burst rates with different time resolutions from seconds to hours. For each time resolution, the burst rate is defined as the number of bursts in each bin divided by the width of the bins.
Then we select the monotonic increasing time structure $\lambda(t)$, according to the selection criteria proposed \textbf{above}. 
The polynomial flare rate function is used to fit the burst rate function:
\begin{equation}
\lambda(t)=\left\{\begin{array}{l}\lambda(t_1)+\left(\lambda_{0}-\lambda(t_1)\right)\left[\left(t-t_{1}\right) /\left(t_{0}-t_{1}\right)\right]^{p} \ \ {\rm for}\ t_{1} \ \leq \ t\ \leq \ t_{0} \\ \lambda(t_2)+\left(\lambda_{0}-\lambda(t_2)\right)\left[\left(t_{2}-t\right) /\left(t_{2}-t_{0}\right)\right]^{p} \ \ \ {\rm for}\ t_{0}\ \leq \ t \ \leq \ t_{2}\end{array}\right.
\end{equation}
where [$t_1$, $t_2$] is the time range of the rise-time structure, and [$t_0$, $\lambda_0$] is an inflection point. There are three free parameters [$\lambda_0$, $t_0$, $p$] in the polynomial flare rate function. 
The rise-time structure of no less than four points is needed since the fit needs more data than free parameters. 
A nonlinear least squares algorithm is used to fit the rise-time structures and the polynomial flare rate function. 
The package \texttt{scipy.optimize.curve\_fit} of Python gives the best-fit value of $p$. 

\begin{figure*}
	\centering
	\includegraphics[width=\textwidth]{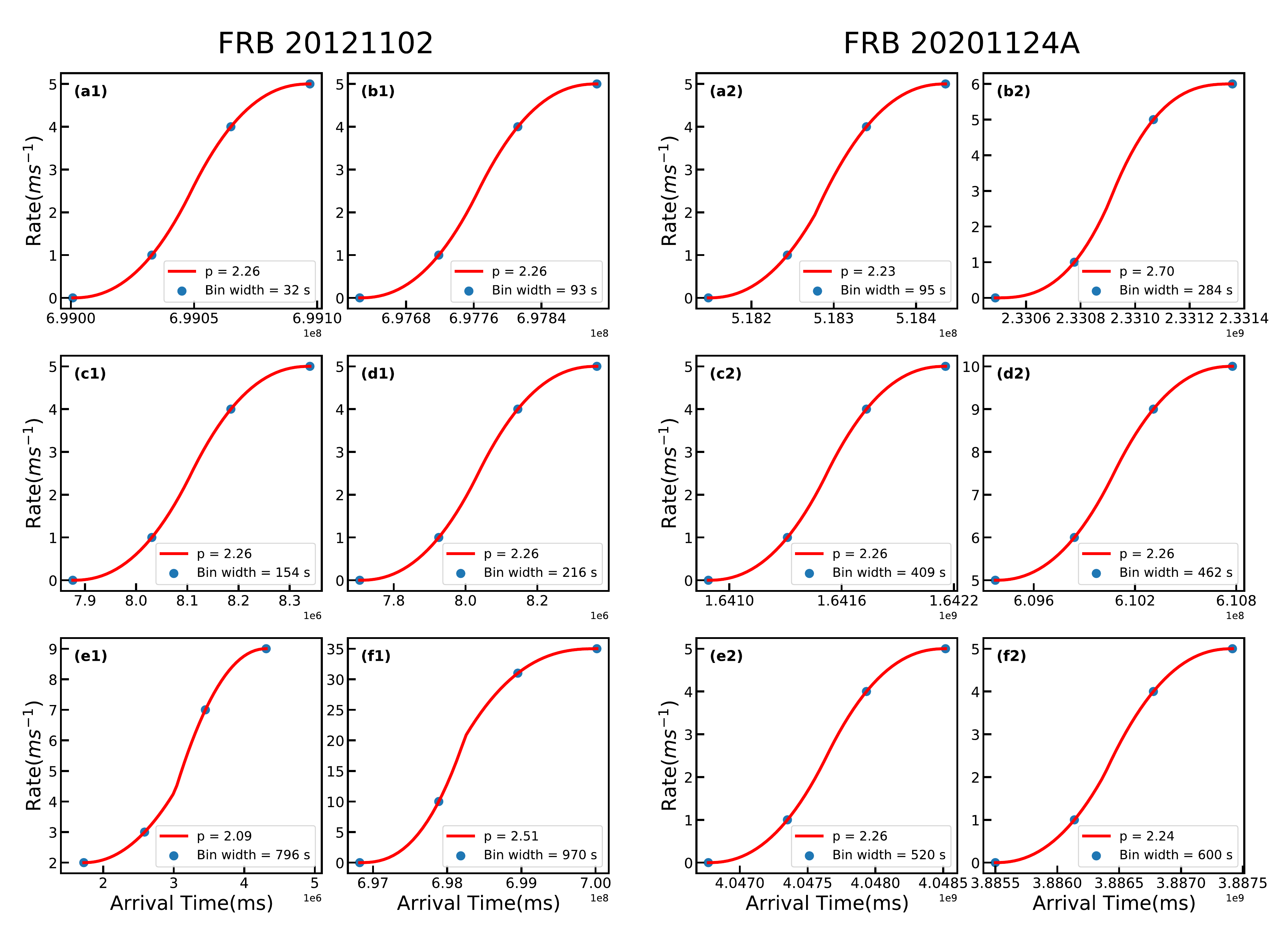}
	\caption{\textbf{Fitted time profiles $\lambda (t)$ of FRB 20121102 and FRB 20201124A.}
		Left panel: The burst rate as a function of the arrival time with different time resolutions. The blue point represents the burst rate. The polynomial curve fitting result is shown as the red solid line.
		The  fitted indices $p$ are given in each sub panel.
		Right panel: Same with the top panel, but for FRB 20201124A. The coherent growth in burst time structures reveals the two repeating FRBs have memory.
	}
	\label{fig:rate}
\end{figure*}

The fitting results of $p$ are given in Table \ref{data:121102} and Table \ref{data:201124} for FRB 20121102 and FRB 20201124A, respectively. A remarkable result is that a large fraction of fitted indices $p$
yield values in a range of $p\geq 2$, which supports that
they are significantly above the linear range $p=1$. So nonlinear physical processes are
responsible for these time structures. Figure \ref{fig:rate} show
some detected time structures with $p>2$ for FRB 20121102 and FRB 20201124A, with time resolution from 32 s to 2,236 s. This demonstrates that
the two repeating FRBs have memory in this timescale. This timescale is well in the range of long waiting times, which can be well modeled by a non-stationary Poisson process. 
This also supports that non-stationary Poisson processes have memory \citep{Aschwanden2021}.
Usually, this coherent growth phase ($\lambda (t)\propto t^p$ with $p\geq 2$) is characteristic for avalanching events that happen in self-organized criticality (SOC) models \citep{Katz1986,Bak1987}.

\subsection{Hurst exponent} \label{sec:Hurst}

\begin{figure*}
	\centering
	\includegraphics[width=\textwidth]{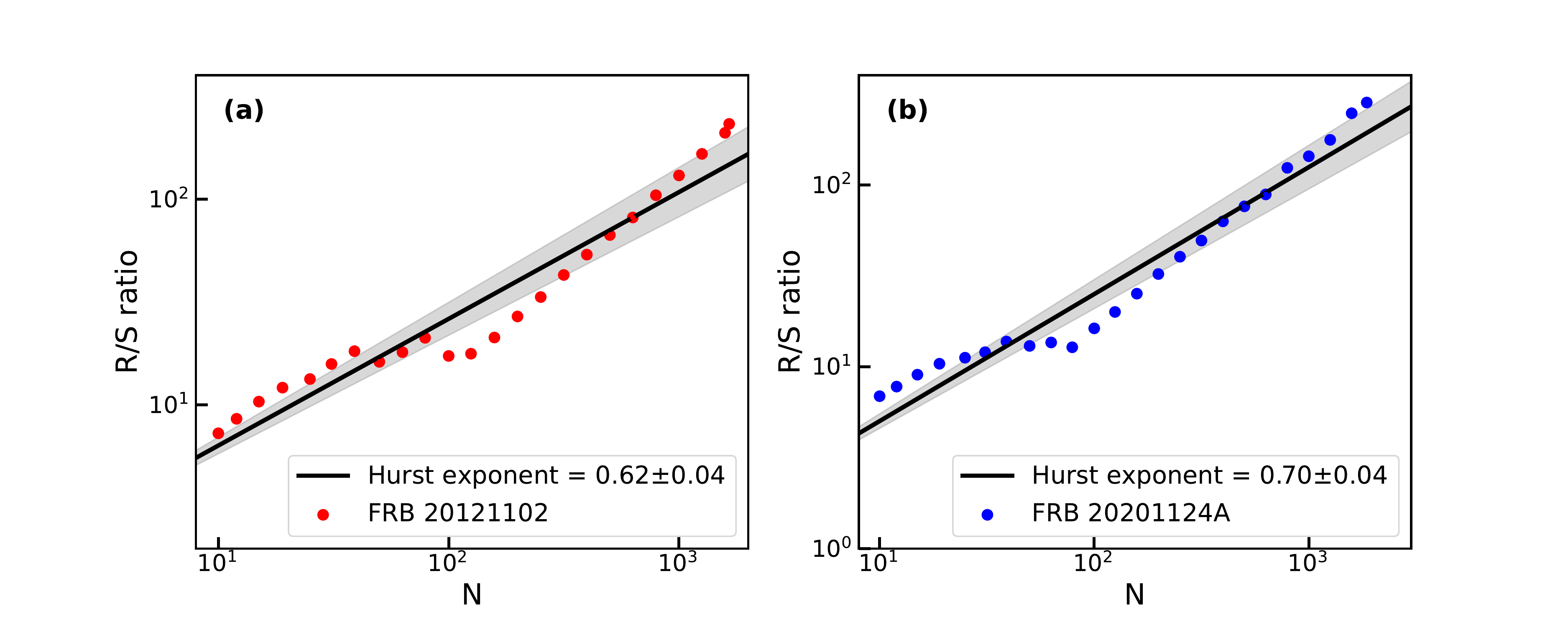}
	\caption{ \textbf{The rescaled range analysis of repeating FRBs.} 
		Panel a. Red scatters correspond to the R/S ratio value of FRB 20121102. The black solid line is the power-law fit and the power-law exponent is the Hurst exponent. 
		The Hurst exponent is 0.62 $\pm$ 0.04 at 1 $\sigma$ confidence level, which is shown as the shaded region.
		Panel b. The blue points and black solid lines are the R/S ratio value and power-law fit of FRB 20201124A, respectively. The Hurst exponent of FRB 20201124A is 0.70 $\pm$ 0.04 at 1 $\sigma$ confidence level, which is shown as shade region.
	}
	\label{fig:Hurst}
\end{figure*}

This subsection defines the Hurst exponent, a useful indicator in describing data and finding whether there is memory in the time series. 
A value of $H=0.5$
indicates that the behavior of the time series corresponds to a stochastic process with no memory. However, $0.5<H\leq 1$ means that the memory is persistent, supporting that a large event is followed by a larger one in the future.
The rescaled range analysis is used to calculate the value of $H$ for repeating FRBs \citep{Mandelbrot1969}. 
For a series of FRB time of arrivals (TOAs) $T = T_1, T_2, ..., T_n$, we remove the unobserved time intervals and set the initial arrival time to zero. 
The average of the processed TOAs $\langle T \rangle$ can be written as
\begin{equation}
\langle T \rangle = \frac{1}{n}\sum_{i = 1}^{n}T_i,
\end{equation}
where $n$ is the total number of bursts during the observation. 
The cumulative value of the difference between the mean at each time and the data, $X(t)$, is defined as
\begin{equation}
X(t) = \sum_{i = 1}^{t}\left(T_i - \langle T \rangle \right), \text{\rm for}\ t = 1, 2, ..., n.
\end{equation}
The self-adjusted range $R(t)$ is
\begin{equation}
R(t) = X_{\rm max}(t) - X_{\rm min}(t), \text{\rm for}\ t = 1, 2, ..., n, 
\end{equation}
where $X_{\rm max}(t)$ and $X_{\rm min}(t)$ are the maximum and the minimum of $X(t)$, respectively.
The standard deviation of $T$ is 
\begin{equation}
S(t) = \left( \frac{1}{t}\sum_{i = 1}^{t}\left(T_i - \langle T \rangle \right)^2 \right)^{\frac{1}{2}}, \text{\rm for}\ t = 1, 2, ..., n.
\end{equation}
The ratio of $R(t)$ and $S(t)$ is related to $t$, and can be written as the function of $t$:
\begin{equation}\label{equa:Hurst}
R/S = C \left({t}\right)^{H}, \text{\rm for}\ t = 1, 2, ..., n, 
\end{equation}
where $C$ is a constant and the exponent $H$ is the so-called Hurst exponent. Thus, the Hurst exponent can be calculated by constructing the $R/S$ value of the arrival times of FRB and fitting the index of Equation (\ref{equa:Hurst}). 
The MCMC analysis is used for calculating the Hurst exponent, and the fitted Hurst exponents are $0.62\pm 0.04$ and $0.70\pm 0.04$ at 1$\sigma$ confidence level for FRB 20121102 and FRB 20201124A, respectively. 
The results of rescaled range analysis are shown in Figure \ref{fig:Hurst}. 
The $H$ values of these two FRBs are in the interval between 0.5 and 1. This supports that they exhibit a persistent behavior, namely long-term memory, which will remain for a long time. The Hurst exponent
confirms the conclusion drawn from the coherent growth in
burst time structures. 

\section{Cellular automaton simulation} \label{sec:simulation}

The differential energy distribution of FRB 20121102 can be fitted by a power-law function, and the power-law indices of bursts detected by different telescopes are almost the same \citep{Wang2019}. 
Analogous characteristic also occurs in numerous astronomical phenomena in different wavelengths \citep{Aschwanden2016}, such as gamma-ray bursts \citep{Wang2013,Yi2016} and solar flares \citep{Aschwanden2016}. The SOC model can reproduce the observed power-law-like distributions, which has been proved by the cellular automaton simulations \citep{Katz1986,Bak1987,Lu1993}. 

\begin{figure*}
	\centering
	\includegraphics[width=\textwidth]{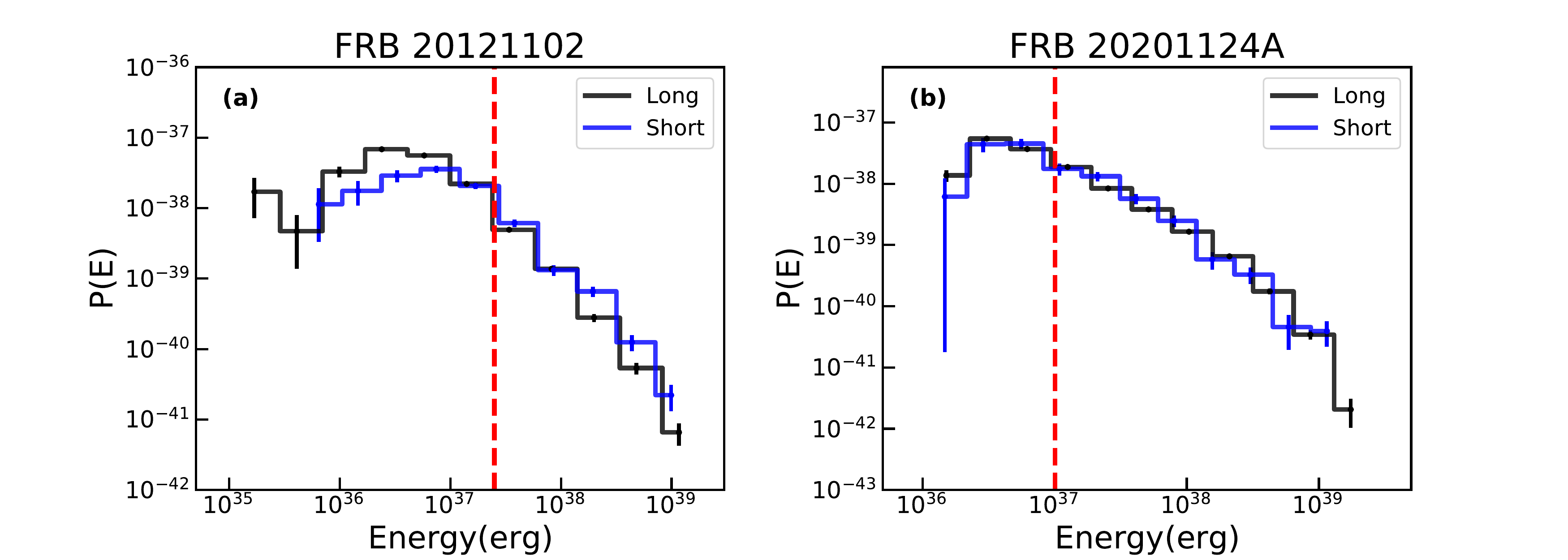}
	\caption{\textbf{The frequency distribution of burst energy for FRB 20121102 and FRB 20201124A.}
		Panel (a). The energy distribution of FRB 20121102. The black solid line with the error bar shows the differential distribution of energy for the ``Long" bursts and the blue solid line for the ``Short" bursts.  The red dashed line at $E = 2.5 \times 10^{37} \ {\rm erg}$ indicates the 90\% detection completeness threshold of FAST. 
		Panel (b). Same with Panel (a), but for FRB 20201124A and the red dashed line shows the 95\% detection completeness threshold of FAST. 
	}
	\label{fig:Energy}
\end{figure*}

As discussed above, the bimodal distributions of the waiting time show that waiting time can be divided into long and short parts. According to this characteristic, we divide the bursts into the ``Long" and the ``Short" parts. 
We present the energy distributions of the ``Long" and the ``Short" parts in Figure \ref{fig:Energy}, respectively. 
The red dashed line of Panel (a) in Figure \ref{fig:Energy} indicates the 90\% detection completeness threshold of FRB 20121102, corresponding to $E_{90} = 2.5 \times 10^{37}$ erg \citep{Li2021}. 
And the red dashed line Panel (b) shows the completeness threshold of energy at the 95\% confidence level of FRB 20201124A, which is calculated according to the completeness threshold of fluence \citep{Xu2022}. 
The power-law tails exist in the energy distributions that exceed the completeness threshold of both the ``Long" and the ``Short" parts. 
Together with the time structures with coherent growth, these characteristics support that repeating FRBs are SOC phenomena \citep{Wang2017,Cheng2020}. However, if external drives
are not correlated (i.e., random driving force), the avalanches (bursts) will occur in a random way, and the probability
distribution function of the waiting times
should be an exponential law, contradicting the observed power-law distributions. In the following, we will use the cellular automaton simulation of the SOC lattice model with time-dependent external drives to explain this discrepancy.

For a dynamic system with external drives input, the system will self-organize to reach the critical state when a critical point is reached \citep{Bak1987}. 
The SOC lattice model with time-dependent drives predicts a power-law tail of the frequency distribution of waiting times for solar flares \citep{Norman2001} and sandpile \citep{Sanchez2002}. 
Here we perform three-dimensional cellular automaton (CA) simulations based on the SOC model with time-dependent external drives.
At present, the origin of FRBs is still unknown. The discovery of FRB 20200428 from the Galactic magnetar indicates that at least some FRBs originate from magnetars \citep{CHIME2020,Bochenek2020}. 
Therefore, we consider the external drive as the perturbation of magnetic field $B$. Although, it can be replaced by other perturbations.
Here we describe the cubic lattice in CA simulations with the magnetic field $B$. 
Regarding a three-dimensional cubic lattice with subscript \{i, j, k\} and the magnetic field of each lattice node is $B_ {\rm i,j,k}$. 
At the boundaries, the magnetic field is assumed to be zero at any time. 
The difference between each cubic lattice and six neighbors can be defined as
\begin{equation}
\Delta B_{\rm i, j, k} = B_{\rm i, j, k} - \frac{1}{6}\sum_{\rm nn}B_{\rm nn},
\end{equation}
where $B_{\rm nn}$ is the difference between a lattice with a neighbor. 
At each time step, random perturbations $\delta B$ are randomly added to the cubic lattice and $\Delta B_{\rm i, j, k}$ is compared with a specified threshold $B_c$. 
If $\Delta B_ {\rm i,j,k}$ is less than $B_c$, the system will keep stability.  
The avalanches occur when $\Delta B_ {\rm i,j,k}$ exceeds the specified threshold $B_c$. 
Then the field is redistributed according to the uniform redistribution rule:
\begin{equation}
B_{\rm i, j, k}' = B_{\rm i, j, k} - \left(\frac{6}{7} \right)s B_c,
\end{equation}
where $s = \Delta B_{\rm i, j, k} / \mid \Delta B_{\rm i, j, k} \mid$. 
After the redistribution of the field, random perturbations are continuously added to the lattice at each time step. The next avalanche will still happen when the threshold is exceeded. 

The form of external drives will have a significant impact on the three-dimensional SOC lattice model \citep{Lu1993,Norman2001}. 
In this study, we consider a random external drive and a time-dependent drive, respectively. 
In the first case, the perturbation $\delta B$ is selected from a random distribution in the range [-0.2, 0.8] \citep{Lu1993}. The situation under which the perturbation $\delta B$ is less than the threshold value $B_c$ needs to be satisfied. 
In the second case, the perturbation $\delta B$ is expressed by a one-dimensional random walk function:
\begin{equation}
\delta B(t) = c \left| \frac{\beta(t)}{\sqrt{d\sqrt{N}}} \right|, 
\end{equation}
where $c$ is a random constant selected from a uniform distribution in the range of [-0.2, 0.8], $\beta (t)$ is the one-dimensional random walk function, the random walk step size $d$ is chosen to be 0.01 and $N$ is the number of random walk steps. 
We record each step of the random walk when a new perturbation is added to the system. Here we set the unit of the waiting time of the consecutive avalanches as an iteration, which presents the time between each successive drive. 

\begin{figure*}
	\centering
	\includegraphics[width=\textwidth]{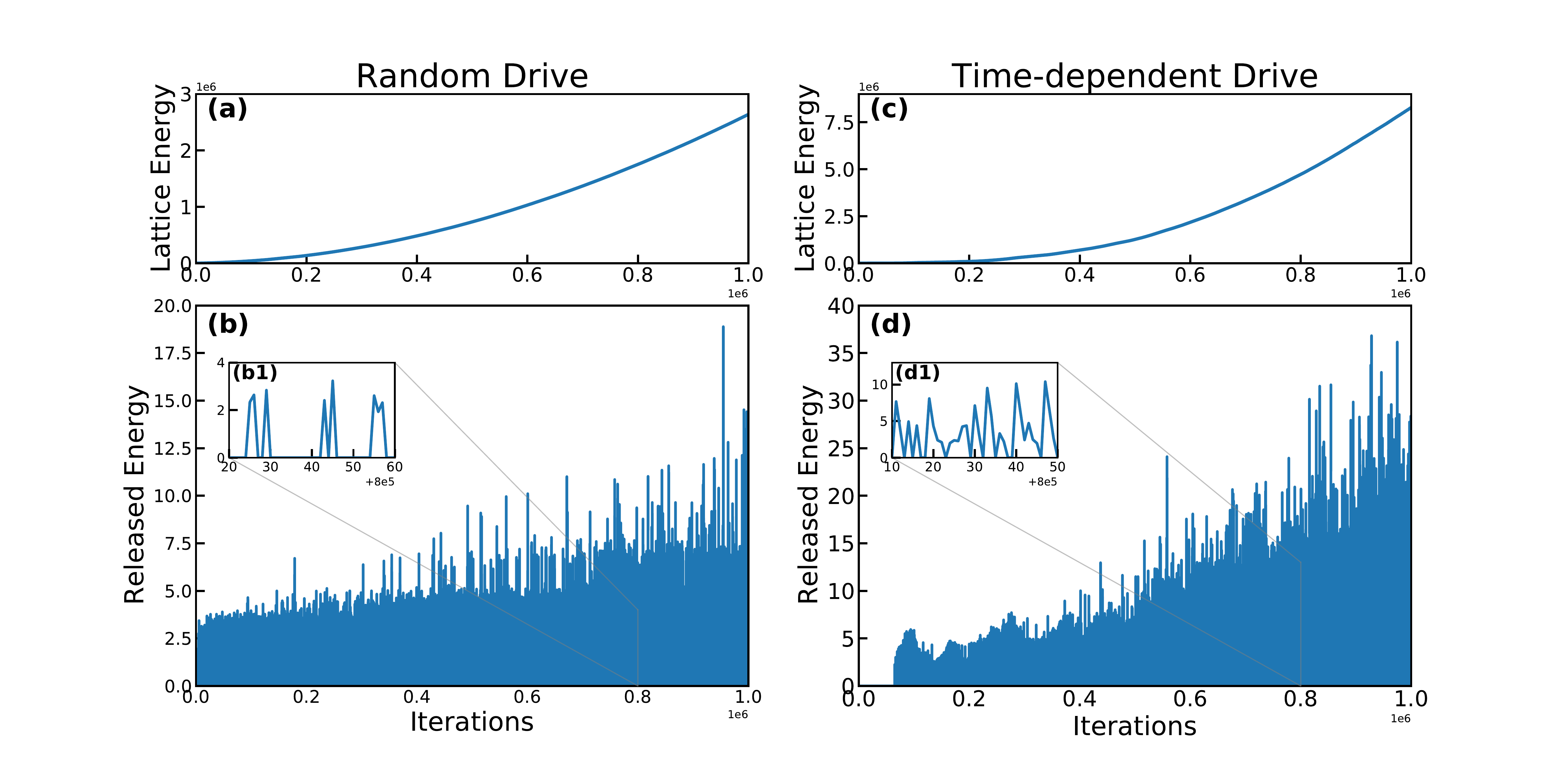}
	\caption{\textbf{Time series of the lattice energy (panels (a) and (c)) and the released energy of simulated avalanches (panels (b) and (d)) in three-dimensional cellular automaton simulation. }
		A $30^3$ lattice with $1\times 10^6$ iterations is applied in this simulation. 
		While two panels on the left represent the SOC process with random drives and two panels on the right represent the time-dependent drives. 
		Panels (b1) and (d1) demonstrate a portion of the time series of released energy, which can identify an individual burst. 
		Simulated waiting time can be derived from the time between two consecutive pulses. 
	}
	\label{fig:Sim1}
\end{figure*}

\begin{figure*}
	\centering
	\includegraphics[width=\textwidth]{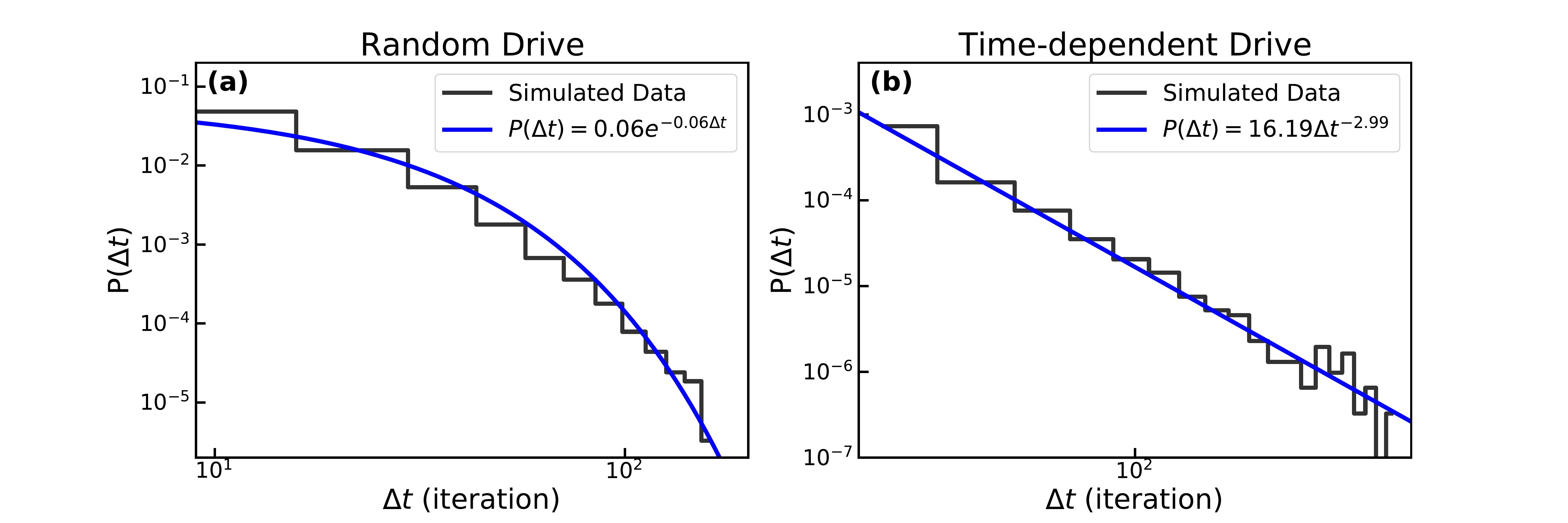}
	\caption{\textbf{Waiting time distributions from cellular automaton simulations of three-dimensional SOC lattice model.} 
		Panel (a). The differential distribution of waiting times for random driving forces. It conforms to a single exponential distribution (blue line), which is characterized by an extreme decrease in long waiting times. 
		Panel (b). The differential distribution of waiting times for the driving force is related to a random walk function. The blue line is the power-law fit for the waiting times. }
	\label{fig:Sim}
\end{figure*}

All simulations are carried out
on a $30^3$ lattice with $10^6$ iterations. The threshold value is set as $B_c = 7$. Here we focus on recording the time between adjacent avalanches and analyze the distribution of waiting times after accumulating enough simulated data. 
The simulation results are shown in Figure \ref{fig:Sim1}.
The panels (a) and (b) of Figure \ref{fig:Sim1} show the results from uniformly distributed drivings. An obvious exponential-tail of waiting time distribution is consistent with the Poisson process, as shown in panel (a) of Figure \ref{fig:Sim}. 
The results of driving force related to the random walk function are shown in panels (c) and (d) of Figure \ref{fig:Sim1}. The waiting time distribution approaches to
power-law function (panel b of Figure \ref{fig:Sim}).
The waiting time distribution of the simulated non-stationary Poisson process is essentially consistent with the waiting time distributions of FRB 20121102 and FRB 20201124A, which supports that the triggers of repeating FRBs are correlated.

\section{Discussion and Conclusions}\label{sec:discussion}

Below, we discuss the physical origin of the correlated
drive. In the framework of the magnetar model, 
zones of magnetic stress are generated in the crust due to magnetic field evolution
and field reconfiguration through
Hall drift and Ohmic dissipation \citep{Ruderman1998}.
When the magnetic stresses exceed a critical threshold, frequent crustal fractures will occur, which can produce repeating FRBs \citep{Suvorov2019,Lu2020,Li2022}. In fracture processes, the crustal strain redistributes through neighbor tectonic interactions, which can trigger subsequent bursts. So tectonic interactions ensure that the triggers of bursts are correlated.
From cellular automaton simulations, power-law distributions of waiting times for crustal failures have been found \citep{Kerin2022}, similar to the waiting times of repeating FRBs.

In summary, from the coherent growths in
burst time structures and the Hurst exponent, we found that repeating FRBs show memory from minutes to an hour, which can be explained by a SOC model.
The power-law distribution of waiting times requires that the bursts are triggered in a correlated way.
One possible trigger mechanism is the crustal failure of a neutron star, favoring emission from the stellar magnetosphere \citep{Suvorov2019,Lu2020,Li2022}.
The coherent growth in the burst time structures and the Hurst exponent of FRB 20121102 and FRB 20201124A support that they exhibit long-term memory. 
These long-term memories suggest that large bursts are more likely to be followed by a larger one, possibly
allowing for the prediction of intense bursts, which is crucial for the searching strategy of repeating FRBs.

\section*{acknowledgements}
We thank the anonymous referee for helpful comments. We acknowledge valuable discussions with Markus J. Aschwanden and W. B. Wang. This work was supported by the National Natural Science Foundation of China (grant Nos. 12273009 and 11833003), the National SKA Program of China (grant Nos. 2022SKA0130100 and 2020SKA0120300),
and the National Key Research and Development Program of China (grant No. 2017YFA0402600). This work made use of data from FAST, a Chinese national mega-science facility built and operated by the National Astronomical Observatories, Chinese Academy of Sciences.

\bibliographystyle{aasjournal}
\bibliography{frb}

\newpage

\clearpage

\clearpage

\clearpage

\clearpage

\clearpage

\clearpage

\begin{table}
   \small
	\begin{center}
		\caption{The reduced $\chi^2$ value for the fitting of waiting time distribution of repeating FRBs with Poisson distribution and non-stationary Poisson distribution. } 
		\begin{tabular}{ccccc}
			\hline
			& \multicolumn{2}{c}{Long waiting time}  &  \multicolumn{2}{c}{Short waiting time}    \\
			& \multicolumn{1}{c}{Non-stationary} & \multicolumn{1}{c}{Poisson}& Non-stationary & Poisson        \\
			\hline
			FRB 20121102  & \multicolumn{1}{c}{1.05}	&   \multicolumn{1}{c}{1.82}&  3.12  &  4.16  \\
			
			FRB 20201124A & \multicolumn{1}{c}{1.00}	&   \multicolumn{1}{c}{7.99} &  6.22  &  4.65  \\
			\hline
		\end{tabular}
		\label{chi2}
		\vspace{0.5cm}
	\end{center}
\end{table}


\begin{table}
	\renewcommand{\arraystretch}{0.5}
	\begin{center}
		\caption{ Nonlinearity index $p$ of burst rate function $\lambda (t)$ as a function of the time Resolution $\Delta t$ and the number of detected events $n_{\rm det}$ of FRB 20121102. }
		\begin{tabular}{ccccc}
			\hline
			Time Resolution & Number of Detected Events & Nonlinearity Index  \\
			$\Delta t\,({\rm s})$      &  $n_{\rm det}$            & $p$           \\
			\hline
			32  &  7   &  [1, 2.25]   \\
			37  &  4 	&  [1, 2.25]   \\
			43  &	7   &  [1, 2.26]   \\
			93  &  4   &  [0.95, 2.26]   \\
			105  & 6   &  [0.55, 2.26]   \\
			120  & 5   &  [0.72, 2.26]   \\
			126  & 4	&  [1, 2.25]   \\
			147  & 3	&  [0.78, 2.71]   \\
			154  & 4	&  [0.61, 3.05]   \\
			177  & 9	&  [0.36, 2.26]   \\
			198  & 8	&  [0.35, 2.59]   \\
			216  & 6	&  [0.58, 2.13]   \\
			230  & 7	&  [0.59, 2]   \\
			407  & 1	&  [2.26]   \\
			681  & 5	&  [0.87, 2.32]   \\
			705  & 4	&  [0.34, 2.32]   \\
			796  & 3	&  [1, 2.09]   \\
			970  & 2	&  [0.4, 2.2]   \\
			1175  & 1	&  [2.67]   \\
			1332  & 1	&  [4.46]   \\
			2236  & 1	&  [2.32]   \\
			\hline
		\end{tabular}
		\label{data:121102}
		\vspace{0.5cm}
	\end{center}
\end{table}

\clearpage

\begin{table}
\renewcommand{\arraystretch}{0.5}
	\begin{center}
		\caption{ Nonlinearity index $p$ of burst rate function $\lambda (t)$ as a function of the time resolution $\Delta t$ and the number of detected events $n_{\rm det}$ of FRB 20201124A. }
		\begin{tabular}{ccccc}
			\hline
			Time Resolution & Number of Detected Events & Nonlinearity Index  \\
			$\Delta t\,({\rm s})$      &  $n_{\rm det}$            & $p$           \\
			\hline
			16  &  2   &   [1, 2.26]  \\
			46  &  4   &   [0.98, 2.25]  \\
			52  &  4   &   [0.42, 2.32]  \\
			59  &  2   &   [2.25, 3.39]  \\
			64  &  3   &   [0.91, 2.16]  \\
			89  &  1   &   [2.97]  \\
			95  &  4   &  [1, 2.26]   \\
			97  &  3 	&  [0.62, 2.25]   \\
			100  &  3   &   [0.58, 2.71]  \\
			112  &  5   &   [1, 2.74]  \\
			139  &  3   &   [1, 2.58]  \\
			284  &	5   &  [0.51, 2.71]   \\
			320  & 7   &  [0.62, 2.71]   \\
			390  & 11   &  [0.54, 2.26]   \\
			409  & 17   &  [0.54, 2.24]   \\
			440  & 11	&  [0.30, 2.21]   \\
			462  & 9	&  [0.30, 2.71]   \\
			499  & 4	&  [1.71, 2.26]   \\
			520  & 6	&  [0.31, 2.25]   \\
			541  & 8	&  [0.31, 2.71]   \\
			561  & 6	&  [0.81, 2.71]   \\
			594  & 5	&  [0.46, 2]   \\
			600  & 6	&  [0.47, 2.26]   \\
			680  & 8	&  [0.37, 3.42]   \\
			685  & 9	&  [0.35, 2.42]   \\
			697  & 10	&  [0.37, 3.42]   \\
			704  & 10	&  [0.37, 3.42]   \\
			724  & 9	&  [0.48, 2.51]   \\
			779  & 4	&  [0.72, 2.51]   \\
			890  & 2	&  [1.71, 2.31]   \\
			1514  & 2	&  [0.41, 2.44]   \\
			\hline
		\end{tabular}
		\label{data:201124}
		\vspace{0.5cm}
	\end{center}
\end{table}

\end{document}